\documentclass[letterpaper,twocolumn,prl,notitlepage,floats,superscriptaddress,amsmath,amssymb]{revtex4-2}

\usepackage[version=3]{mhchem} 
\usepackage{bm}
\usepackage[utf8]{inputenc}
\usepackage[T1]{fontenc}
\usepackage{graphicx}
\usepackage{upgreek}
\usepackage{color}
\usepackage{ulem}
\usepackage{hyperref} 
\hypersetup{colorlinks, citecolor=blue, filecolor=blue, linkcolor=blue, urlcolor=blue}
\usepackage{sidecap}
\usepackage{amssymb}
\usepackage[caption=false]{subfig}
\usepackage{xcolor}
\usepackage{float}

\def \FUW{Faculty of Physics, University of Warsaw, 02-093 Warsaw, Poland}
\def \Sapienza{Physics Department, Sapienza University of Rome, 00185 Rome, Italy}

\def \PWR{Institute of Theoretical Physics, Wrocław University of Science and Technology, 50-370 Wrocław, Poland}

\begin{document}
	
	\title{Layer-Dependent Vibrational and Optical Properties of \ce{Mo_{0.58}W_{0.42}Se_2} Alloy}
	
	\author{Szymon Socha}
	\email{s.socha3@student.uw.edu.pl}
	\affiliation{\FUW}
	\author{Tomasz Woźniak}
	\affiliation{\PWR,\FUW}
	\author{Elena Blundo}
	\affiliation{\Sapienza}
	\author{Małgorzata Brzoska}
	\affiliation{\FUW}
	\author{Grzegorz Krasucki}
	\affiliation{\FUW}
	\author{Piotr~Wróbel}
	\affiliation{\FUW}
	\author{Antonio Polimeni}
	\affiliation{\Sapienza}
	\author{Adam Babiński}
	\affiliation{\FUW}
	\author{Maciej R. Molas}
	\email{maciej.molas@fuw.edu.pl}
	\affiliation{\FUW}
	\author{Katarzyna Olkowska-Pucko}
	\email{katarzyna.olkowska-pucko@fuw.edu.pl}
	\affiliation{\FUW}
	
	\begin{abstract}
		Semiconducting Mo$_x$W$_{1-x}$Se$_2$ alloys provide a versatile platform for tailoring the optical properties of two-dimensional materials through both composition and layer thickness. 
		Here, we systematically investigate mechanically exfoliated Mo$_{0.58}$W$_{0.42}$Se$_2$ flakes ranging from monolayer (1L) to nine layers by combining Raman scattering (RS), photoluminescence (PL), reflectance contrast (RC) spectroscopy, and first-principles phonon calculations. 
		Thirteen RS peaks are identified, including the low-frequency interlayer shear mode, whose thickness dependence is well described by a linear-chain model, yielding an interlayer force constant of $K_s=(2.996\pm0.015)\times10^{19}$ N m$^{-3}$. 
		PL measurements reveal a crossover from the direct-bandgap 1L to indirect-bandgap multilayers. 
		The thickness evolution of the indirect optical transition is quantitatively reproduced using a quantum-confinement model, yielding an out-of-plane reduced effective mass of $\mu_\perp=0.75 m_0$. 
		RC spectroscopy reveals four excitonic resonances. 
		While the A and B excitons associated with the $K^\pm$ valleys remain nearly independent of layer thickness, the higher-energy C and D resonances originating from the band-nesting regions exhibit pronounced redshifts, reflecting substantial thickness-induced modifications of the electronic band structure. 
		These results establish comprehensive spectroscopic fingerprints of flake thickness, interlayer coupling, and electronic structure in Mo$_x$W$_{1-x}$Se$_2$ alloys and provide a reliable, non-destructive framework for their optical characterization.
		
	\end{abstract}
	
	\maketitle

	\section{Introduction}
	
	Semiconducting transition metal dichalcogenides (TMDs), including MoSe$_2$, WSe$_2$, MoS$_2$, and WS$_2$, constitute a family of two-dimensional (2D) materials with remarkable optical and electronic properties, making them attractive for both fundamental studies and optoelectronic applications \cite{Wang2012,Mak2016}. When thinned down to a monolayer (1L), these materials undergo a transition from an indirect to a direct band gap, resulting in intense excitonic emission and pronounced optical resonances that can be efficiently investigated by photoluminescence (PL) and reflectance contrast (RC) spectroscopy \cite{Splendiani2010,Mak2010,Arora2015W,Arora2015Mo,Koperski2017,Cadiz2017,Molas2017}. 
	Consequently, increasing the number of layers leads to strong PL quenching and the emergence of indirect optical transitions as the electronic band structure gradually approaches the bulk limit \cite{Padilha2014,Xie2015,Molas2017,Mak2018,Catanzaro2024}.
	Although binary TMDs already cover a broad range of optical properties, the accessible spectral range remains inherently limited.
	For example, the A-exciton energies of monolayer MoSe$_2$ and WSe$_2$ are approximately 1.6 and 1.7~eV, respectively \cite{Tonndorf2013,Arora2015W,Arora2015Mo}, leaving the intermediate spectral region inaccessible using the parent compounds alone. 
	Ternary alloys such as Mo$_x$W$_{1-x}$Se$_2$ overcome this limitation by enabling continuous tuning of the electronic band structure through compositional engineering \cite{Zhang2014,Xie2015,Meng2019,Kopaczek2021,Nugera2022}. 
	Consequently, their optical properties can be tailored over a wide spectral range, making these materials promising for optoelectronic and valleytronic applications \cite{Pal2023}.
	
	Besides excitonic transitions, lattice vibrations provide complementary information on the structural properties and interlayer interactions in TMDs. 
	Raman scattering (RS) is a powerful, non-destructive technique for probing these vibrational excitations and their evolution with flake thickness \cite{Lee2010,Zhao2013,Zhang2016}. 
	In particular, both high-frequency intralayer phonon modes and low-frequency interlayer shear modes exhibit systematic frequency shifts with increasing layer number, providing direct insight into lattice dynamics and interlayer coupling \cite{Lee2010,Zhao2013,OBrien2015,Grzeszczyk2016,kipczak2020optical}. 
	Alloying further modifies the vibrational properties of TMDs owing to the different atomic masses and local bonding environments of the constituent elements, leading to changes in phonon frequencies, linewidths, and intensities relative to the parent compounds \cite{Zhang2014,Qingan2022,Proupin2024,Jung2024,Kolesov2026}.
	Recent studies have investigated the composition-dependent optical response of Mo$_x$W$_{1-x}$Se$_2$ alloys and, more recently, their thickness-dependent Raman spectra using the modified random element isodisplacement (MREI) model \cite{Zhang2014,Kolesov2026}. 
	However, a comprehensive investigation combining RS, PL, and RC spectroscopy with first-principles phonon calculations over a broad range of layer thicknesses is still lacking.
	In particular, the low-frequency interlayer shear mode has not yet been systematically investigated in Mo$_{1-x}$W$_x$Se$_2$ alloys despite its importance for understanding interlayer coupling.
	
	\begin{figure*}[t!]
		\centering
		\includegraphics[width=1\linewidth]{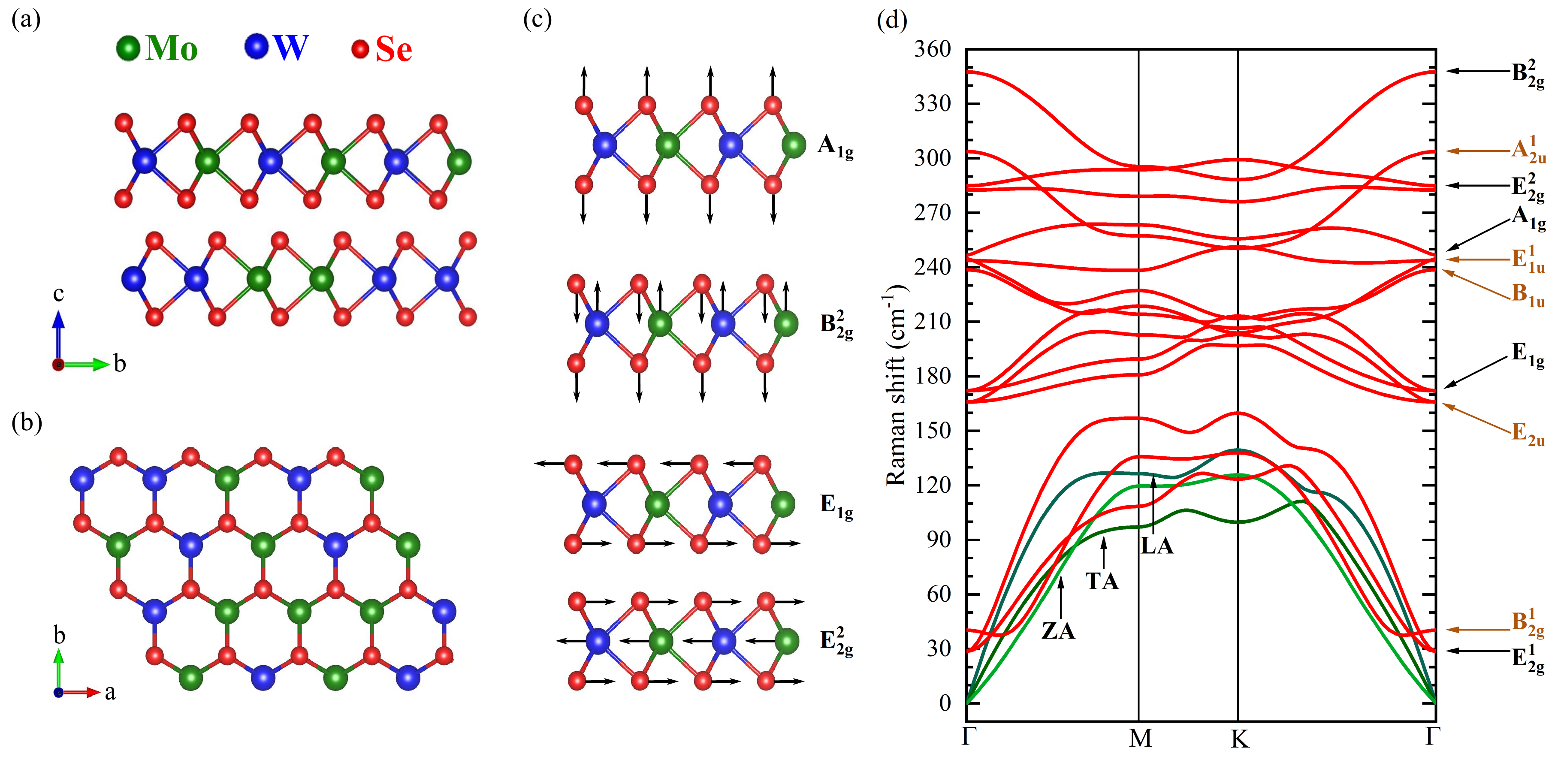}
		\caption{Crystal structure of the hexagonal 2H Mo$_{0.58}$W$_{0.42}$Se$_2$ alloy: (a) side view along the $a$ axis showing the bilayer (2L) structure and (b) top view along the $c$ axis of a monolayer (1L).(c) Schematic representation of vibrational modes at the $\Gamma$ point of the first BZ in monolayer MoWSe$_2$. (d) Phonon dispersion of bulk Mo$_{0.5}$W$_{0.5}$Se$_2$, calculated using density functional theory (DFT). The phonon branches are labeled according to the irreducible representations of the $\mathrm{D}^4_{6h}$ point group at the $\Gamma$ point. Phonon modes identified in the experimental Raman spectra are indicated by black labels, whereas those not observed are shown in brown. 
		}
		\label{fig:structure}
	\end{figure*}
	
	In this work, we combine RS, PL, and RC spectroscopy with first-principles phonon dispersion calculations to systematically investigate the thickness-dependent optical and vibrational properties of mechanically exfoliated Mo$_{0.58}$W$_{0.42}$Se$_2$ flakes ranging from 1L to nine layers (9L).
	The calculated phonon dispersion enables reliable assignment of both first- and second-order Raman features, while the observation of the low-frequency interlayer shear mode allows to quantify the interlayer coupling strength within the linear-chain model.
	Together, these results reveal the evolution of lattice dynamics and electronic structure with increasing layer thickness and establish characteristic spectroscopic fingerprints of MoWSe$_2$ alloys.
	
	\subsection{Crystal structure and phonon dispersion}
	
	The investigated Mo$_{0.58}$W$_{0.42}$Se$_2$ alloy crystallizes in the hexagonal 2H polytype~\cite{Zhang2014,Wang2015}, as schematically illustrated in Figs.~\ref{fig:structure}(a) and (b).
	Each monolayer consists of a transition-metal plane with randomly distributed Mo and W atoms sandwiched between two Se atomic planes, thereby preserving the characteristic layered MX$_2$ crystal structure of the parent compounds, MoSe$_2$ and WSe$_2$~\cite{Patel2019,Taank2024,Gao2025}.
	The atoms within each layer are connected by strong covalent bonds, whereas adjacent layers interact through weak van der Waals forces.
	Although the alloy exhibits atomic-scale chemical disorder arising from the random distribution of Mo and W atoms, its average composition remains homogeneous on the macroscopic scale, as confirmed by energy-dispersive X-ray spectroscopy (EDX), which yields a Mo concentration of $x=0.58$ [see Supplementary Information (SI) for details].
	The random distribution of Mo and W atoms modifies the local chemical environment and, consequently, the lattice dynamics compared with those of the parent compounds MoSe$_2$ and WSe$_2$.
	Despite the differences in the atomic masses and bonding characteristics of Mo and W, these modifications are expected to manifest primarily as shifts and broadening of the phonon modes rather than qualitative changes in the phonon dispersion~\cite{Zhang2014,Qingan2022,Proupin2024,Jung2024,Kolesov2026}.
	
	Bulk 2H-MoSe$_2$ and WSe$_2$ belong to the $D_{6h}$ point group~\cite{Wilson1969,Verble1970}. For the ideal 2H structure, the primitive unit cell contains two MX$_2$ layers (six atoms), the lattice vibrations comprise eighteen phonon branches, including three acoustic and fifteen optical modes. 
	The corresponding vibrational representation at the $\Gamma$ point of the Brillouin zone (BZ) is
	\begin{equation}
		\Gamma \equiv
		\mathrm{A_{1g}\oplus2A_{2u}\oplus2B_{2g}\oplus B_{1u} \\
			\oplus E_{1g}\oplus2E_{1u}\oplus2E_{2g}\oplus E_{2u}}.
	\end{equation}
	For pristine few-layer TMD crystals, the point-group symmetry is $D_{3h}$ for an odd number of layers and $D_{3d}$ for an even number of layers. In random Mo$_x$W$_{1-x}$Se$_2$ alloys, the random occupation of the transition-metal sublattice lowers the crystal symmetry. The monolayer retains $D_{3h}$ symmetry, whereas multilayer structures are described by the $D_6$ point group.
	
	Nevertheless, following the convention commonly adopted for layered TMDs~\cite{Zhang2015_raman}, the experimentally observed Raman modes are labeled throughout this work using the bulk notation.
	The atomic displacement patterns of the Raman-active phonons~~\cite{Lee2010,Molina2011,Zhao2013,Yamamoto2014} are schematically illustrated in Fig.~\ref{fig:structure}(c).
	The A$_{\mathrm{1g}}$ and B$_{\mathrm{2g}}^{1}$ modes correspond to out-of-plane atomic vibrations, whereas the E$_{\mathrm{1g}}$ and E$_{\mathrm{2g}}^{2}$ modes involve in-plane atomic vibrations.
	
	To establish a theoretical framework for assigning the experimentally observed Raman features, we performed first-principles calculations of the phonon dispersion.
	First-principles modeling of the experimentally investigated random alloy with the EDX-determined composition Mo$_{0.58}$W$_{0.42}$Se$_2$ would require very large supercells to accurately reproduce the random occupation of the transition-metal sublattice.
	Since such calculations are computationally demanding and are not expected to qualitatively alter the main features of the phonon spectrum, we employed an ordered bulk Mo$_{0.5}$W$_{0.5}$Se$_2$ model, in which Mo and W atoms periodically occupy the transition-metal sublattice within the primitive cell.
	Although the modeled composition differs slightly from the experimental one, the resulting phonon dispersion captures the vibrational properties relevant to the present study and provides a reliable reference for assigning the experimentally observed Raman modes.
	
	Figure~\ref{fig:structure}(d) presents the phonon dispersion of bulk Mo$_{0.5}$W$_{0.5}$Se$_2$ calculated within density functional theory (DFT).
	The low-frequency optical region contains the interlayer shear and layer-breathing modes, which arise from the relative motion of adjacent MX$_2$ layers.
	In the present work, however, only the interlayer shear mode is experimentally resolved and analyzed.
	The calculated phonon energies and eigenvectors are subsequently used to identify the symmetries and atomic displacement patterns associated with the Raman features observed experimentally.
	The experimentally observed modes are indicated by black labels in Fig.~\ref{fig:structure}(d), whereas the remaining phonon branches, including Raman-inactive and infrared-active modes, are shown in brown.


	\subsection{Identification and assignment of Raman modes}
	
	\begin{figure}[h!]
		\centering
		\includegraphics[width=1\linewidth]{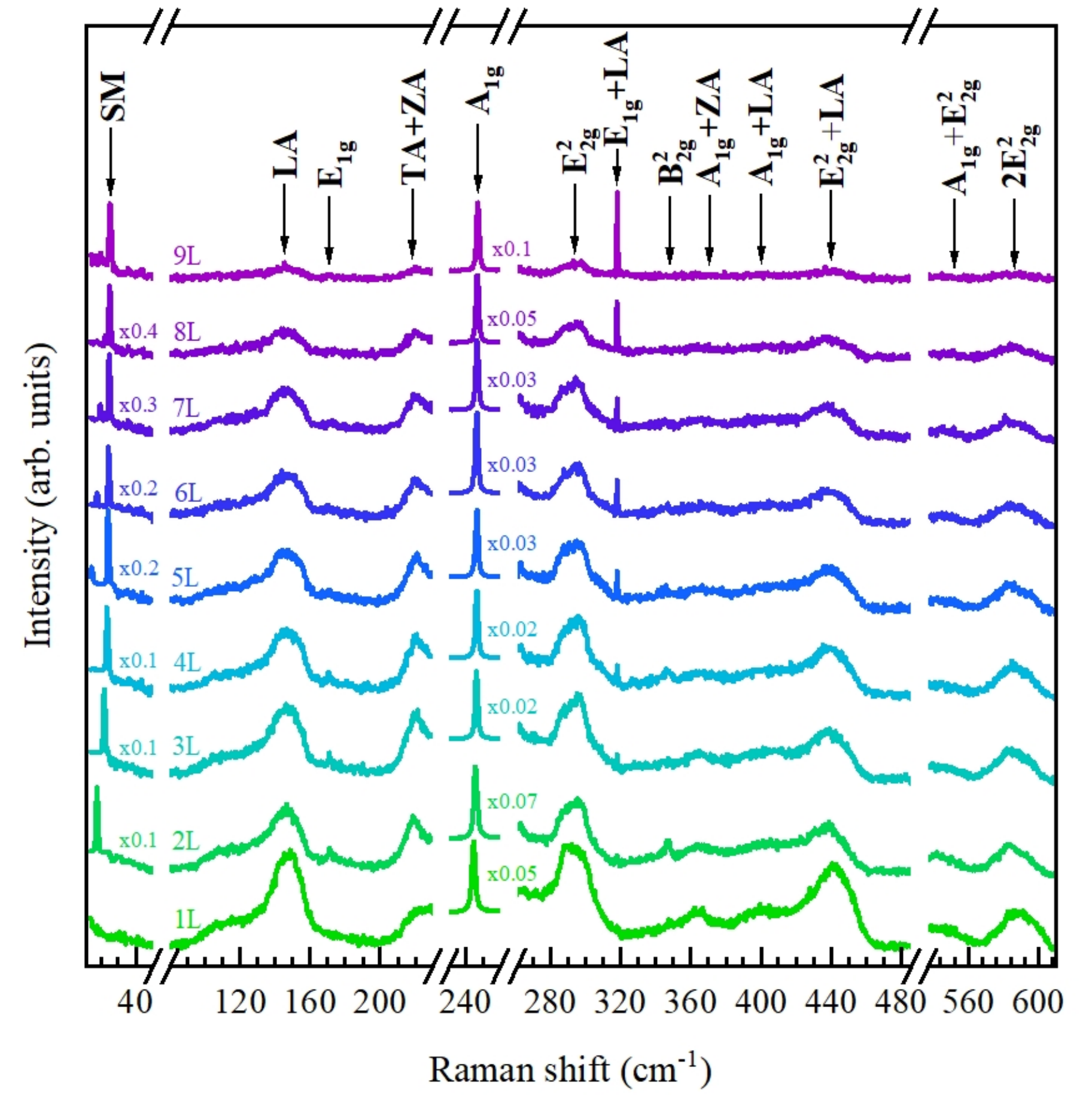}
		\caption{Room-temperature Stokes Raman spectra of mechanically exfoliated Mo$_{0.58}$W$_{0.42}$Se$_2$ flakes with thicknesses ranging from 1L to 9L. The observed Raman modes are labeled using the conventional bulk notation, while their detailed assignments to the corresponding high-symmetry points of the Brillouin zone are discussed in the text and summarized in the Supplementary Information. The spectra are vertically shifted for clarity. The low-frequency and A$_{1g}$ spectral regions were independently scaled with respect to the shear and A$_{1g}$ mode intensities, respectively, with the corresponding scaling factors indicated in the figure.}
		\label{fig:Dyspersion_raman}
	\end{figure}
	
	Figure~\ref{fig:Dyspersion_raman} presents the room-temperature Stokes Raman spectra of mechanically exfoliated Mo$_{0.58}$W$_{0.42}$Se$_2$ flakes with thicknesses ranging from 1L to 9L.
	Optical microscopy images of the investigated flakes, together with AFM topography maps and the corresponding thickness determination, are provided in the SI.
	We begin the mode assignment with the 9L flake, whose Raman spectrum is closest to the bulk limit and therefore allows direct comparison with the calculated phonon dispersion shown in Fig.~\ref{fig:structure}(d).
	The spectrum is dominated by the intense first-order $\mathrm{A_{1g}}$ phonon from the $\Gamma$ point at 247~cm$^{-1}$. 
	In addition, the first-order $\Gamma$-point phonons $\mathrm{E_{1g}}$ and $\mathrm{E^{2}_{2g}}$ are observed at 170 and 294~cm$^{-1}$, respectively.
	These experimental energies are in good agreement with the DFT-calculated values of 247, 172 and 285~cm$^{-1}$ for the $\mathrm{E_{1g}}$, $\mathrm{A_{1g}}$, and $\mathrm{E^{2}_{2g}}$ modes, respectively, supporting the reliability of the calculated phonon dispersion.
	Although the $\mathrm{E_{1g}}$ mode is already visible in the 9L spectrum, it becomes more clearly resolved in thinner flakes. A pronounced thickness dependence of the visibility and relative intensity of this mode has also been reported for few-layer MoTe$_2$, MoSe$_2$, and WSe$_2$~\cite{Goldstein2016,Kim2016,Blaga2024}.
	A weak Raman feature at approximately 348~cm$^{-1}$ is assigned to the $\Gamma$-point $\mathrm{B^{2}_{2g}}$ phonon, in excellent agreement with DFT calculations.
	This mode is Raman inactive in bulk crystals due to the $D_{6h}$ symmetry but becomes Raman active in few-layer crystals as a consequence of symmetry reduction, consistent with previous observations in MoTe$_2$ and related layered materials~\cite{Yamamoto2014,Grzeszczyk2016}.
	
	The low-frequency interlayer shear modes (SM) are also clearly resolved in flakes with thicknesses ranging from 2L to 9L.
	These modes originate from the relative in-plane displacement of adjacent MX$_2$ layers and evolve into the Raman-active $\mathrm{E^{1}_{2g}}$ phonon in the bulk limit.
	Throughout this work, we adopt the conventional shear-mode (SM) notation commonly used for layered TMDs~\cite{Zhao2013,OBrien2015,Zhang2015_raman,Grzeszczyk2016}.
	
	Since the Raman measurements were performed under resonant excitation conditions, the spectra contain not only first-order Raman-active phonons at the $\Gamma$ point but also features involving phonons with finite wave vectors. The calculated phonon density of states, provided in the Supplementary Information, further supports the assignment of these features.
	Under resonant excitation, the Raman momentum selection rule is relaxed, allowing phonons away from the BZ center, particularly those near the $M$ and $K$ point, to contribute to the scattering process~\cite{golasa2015,Soubelet2016,Gontijo2019,Tan2021}. Previous studies favored phonons near the $M$ point based on momentum conservation and resonant optical absorption~\cite{golasa2014resonant,Guo2015}. However, double-resonance calculations indicate that phonons near both $K$ and $M$ may contribute~\cite{Carvalho2017}. Therefore, in the following, the observed features are assigned to specific phonon branches without distinguishing between the $M$ and $K$ points in the BZ.
	Accordingly the lowest-energy resonant feature, observed at 145~cm$^{-1}$, is assigned to the longitudinal acoustic LA phonon.
	This assignment agrees well with both the calculated phonon dispersion and previous reports for MoSe$_2$, WSe$_2$, and related Mo$_{x}$W$_{1-x}$Se$_2$ alloys~\cite{Zhang2014,Zhang2015_raman,Shinde2021}.
	
	Resonant excitation also activates several higher-order Raman features arising from multiphonon scattering processes.
	The Raman peak at 218~cm$^{-1}$ is assigned to the $\mathrm{TA+ZA}$ combination mode.
	In the higher-frequency range (300--590~cm$^{-1}$), additional multiphonon features are identified and assigned to $\mathrm{E_{1g}+LA}$ (317~cm$^{-1}$), $\mathrm{A_{1g}+ZA}$ (370~cm$^{-1}$), $\mathrm{A_{1g}+LA}$ (400~cm$^{-1}$), $\mathrm{E^{2}_{2g}+LA}$ (440~cm$^{-1}$), $\mathrm{A_{1g}+E^{2}_{2g}}$ (551~cm$^{-1}$), and $2\mathrm{E^{2}_{2g}}$ (586~cm$^{-1}$).
	These assignments are supported by the calculated phonon dispersion and are consistent with previous reports for binary TMDs and related Mo$_{x}$W$_{1-x}$Se$_2$ alloys~\cite{Zhang2014,Zhang2015_raman,Shinde2021}.
	
	Comparison of the experimental Raman spectra with the calculated phonon dispersion, together with previous reports for MoSe$_2$, WSe$_2$, and related Mo$_{x}$W$_{1-x}$Se$_2$ alloys, enables thirteen Raman features to be identified and assigned to the corresponding phonon modes.
	The complete mode assignment, together with a comparison of the calculated and experimental phonon energies, is summarized in Table~S1 of the SI.
	The overall agreement between theory and experiment demonstrates that, under resonant excitation, the Raman response of Mo$_{0.58}$W$_{0.42}$Se$_2$ contains contributions from both zone-center and finite-wave-vector phonons, highlighting the important role of exciton-mediated phonon activation throughout the BZ.
	
	\begin{figure*}[htpb]
		\centering
		\includegraphics[width=1\linewidth]{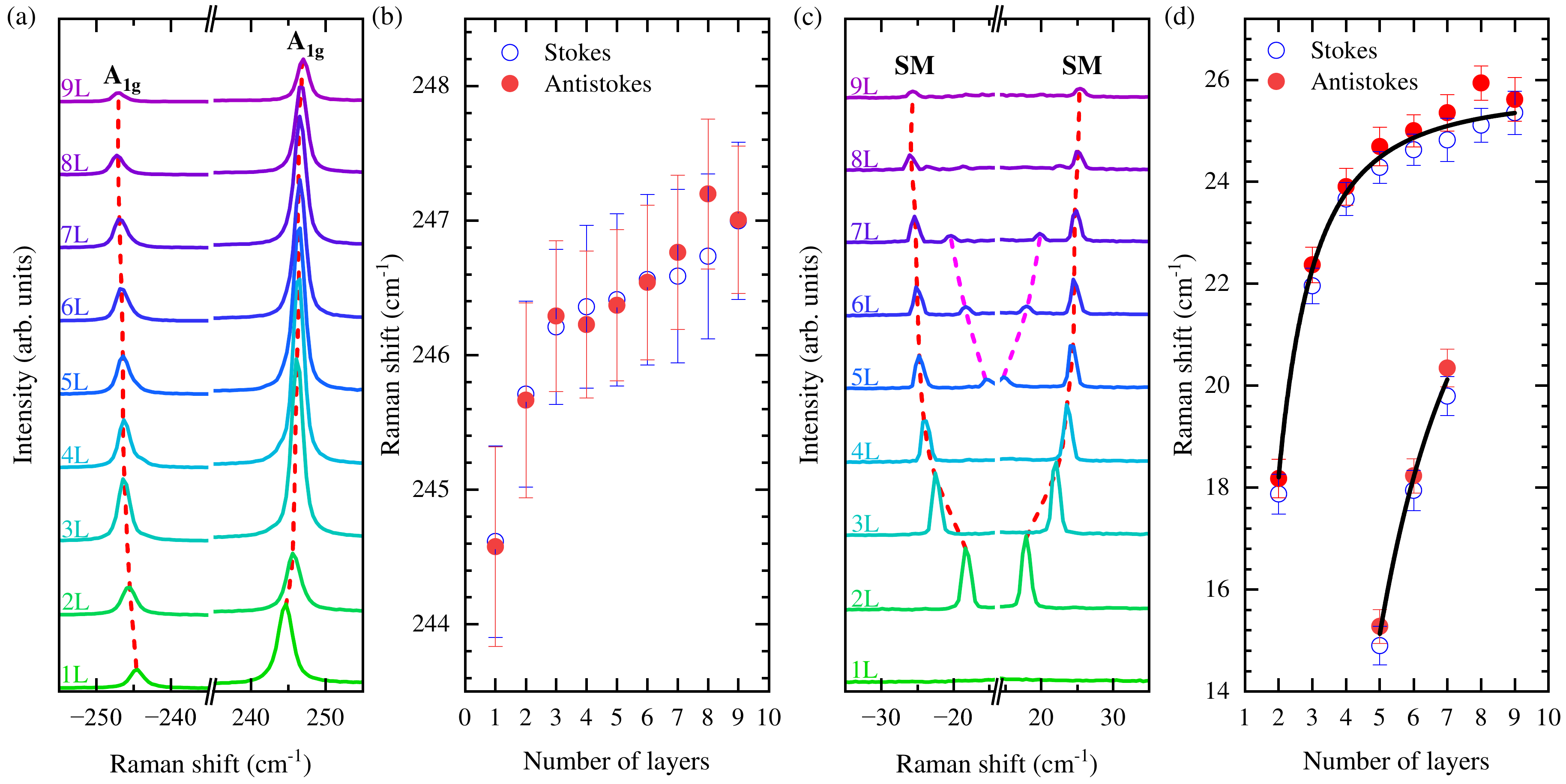}
		\caption{(a) Stokes and anti-Stokes Raman spectra in the vicinity of the $\mathrm{A}_{1g}$ mode. The dashed lines are guides to the eye illustrating the evolution of the phonon energy with layer thickness. (b) Thickness dependence of the $\mathrm{A}_{1g}$ phonon energy. Error bars correspond to one-third of the full with of the half maximum (FWHM) obtained from the Lorentzian fits. (c) Stokes and anti-Stokes Raman spectra in the vicinity of the interlayer shear mode (SM). The dashed lines are guides to the eye illustrating the evolution of the phonon energy with layer thickness. (d) Thickness dependence of the shear-mode energy. Error bars correspond to one-third of the FWHM obtained from the Lorentzian fits. The solid line represents a fit to the linear-chain model.}
		\label{fig:RS}
	\end{figure*}
	
	The identified phonon modes provide the basis for the thickness-dependent Raman analysis presented below. 
	Particular attention is devoted to the interlayer shear mode and the $\mathrm{A_{1g}}$ phonon, whose frequencies are highly sensitive to layer thickness and interlayer coupling in layered TMDs~\cite{Lee2010,Zhao2013,OBrien2015,Grzeszczyk2016}.
	
	\subsection{Thickness-dependent evolution of the \ce{A_{1g}} and shear modes}
	
	Figure~\ref{fig:RS}(a) presents the thickness evolution of the Stokes and anti-Stokes Raman spectra in the vicinity of the $\mathrm{A_{1g}}$ phonon for Mo$_{0.58}$W$_{0.42}$Se$_2$ flakes. 
	The phonon energies were extracted by fitting the Raman peaks with Lorentzian functions, and the resulting values are summarized in Fig.~\ref{fig:RS}(b). 
	For clarity, the absolute values of the anti-Stokes Raman shifts are plotted together with the Stokes data. 
	The excellent agreement between the independently determined Stokes and anti-Stokes energies confirms the reliability of the fitting procedure. 
	The $\mathrm{A_{1g}}$ mode exhibits a systematic blueshift with increasing layer thickness, from 244.5~cm$^{-1}$ in the monolayer to 247.0~cm$^{-1}$ in the 9L flake, corresponding to a total shift of approximately 2.5~cm$^{-1}$. 
	This behavior is consistent with previous reports for binary TMDs, including MoTe$_2$~\cite{Yamamoto2014,Froehlicher2015,Grzeszczyk2016}, MoS$_2$~\cite{Lee2010}, MoSe$_2$~\cite{Tonndorf2013, Zhang2019_2}, and WS$_2$/WSe$_2$~\cite{Zhao2013}. 
	The gradual saturation of the phonon energy for thicker flakes reflects the approach to the bulk limit, while the strong intensity and monotonic thickness dependence of the $\mathrm{A_{1g}}$ mode make it a reliable spectroscopic fingerprint of the layer number in Mo$_{0.58}$W$_{0.42}$Se$_2$.

	We now focus on the low-frequency Raman spectra shown in Fig.~\ref{fig:RS}(c), which reveal the interlayer shear modes. 
	The spectra were measured in both Stokes and anti-Stokes configurations over the Raman-shift range of 14--35~cm$^{-1}$. 
	As expected for an interlayer vibration, the shear mode is absent in 1L and appears only in multilayer flakes. 
	The energy of the main shear-mode branch increases monotonically with increasing layer thickness, while an additional branch becomes resolved for flakes from 5L to 7L, consistent with previous observations in TMDs~\cite{Liang2017}. 
	The extracted mode energies are summarized in Fig.~\ref{fig:RS}(d). 
	The main branch shifts from approximately 18~cm$^{-1}$ in 2L to 26~cm$^{-1}$ in 9L, whereas the second branch increases from about 15~cm$^{-1}$ in 5L to approximately 20~cm$^{-1}$ in 7L. 
	This pronounced thickness dependence makes the shear modes sensitive spectroscopic fingerprints of the layer number and direct probes of interlayer coupling~\cite{Zhao2013,OBrien2015,Grzeszczyk2016,Pizzi2021}. For comparison, the shear-mode energies in 2L MoSe$_2$ and WSe$_2$ are approximately 18 and 17~cm$^{-1}$, respectively~\cite{OBrien2015}. 
	The value measured for 2L Mo$_{0.58}$W$_{0.42}$Se$_2$ therefore lies between those of the parent compounds, as expected for an alloy of intermediate composition.
	
	\begin{figure*}[htpb]
		\subfloat{}
		\centering
		\includegraphics[width=1\linewidth]{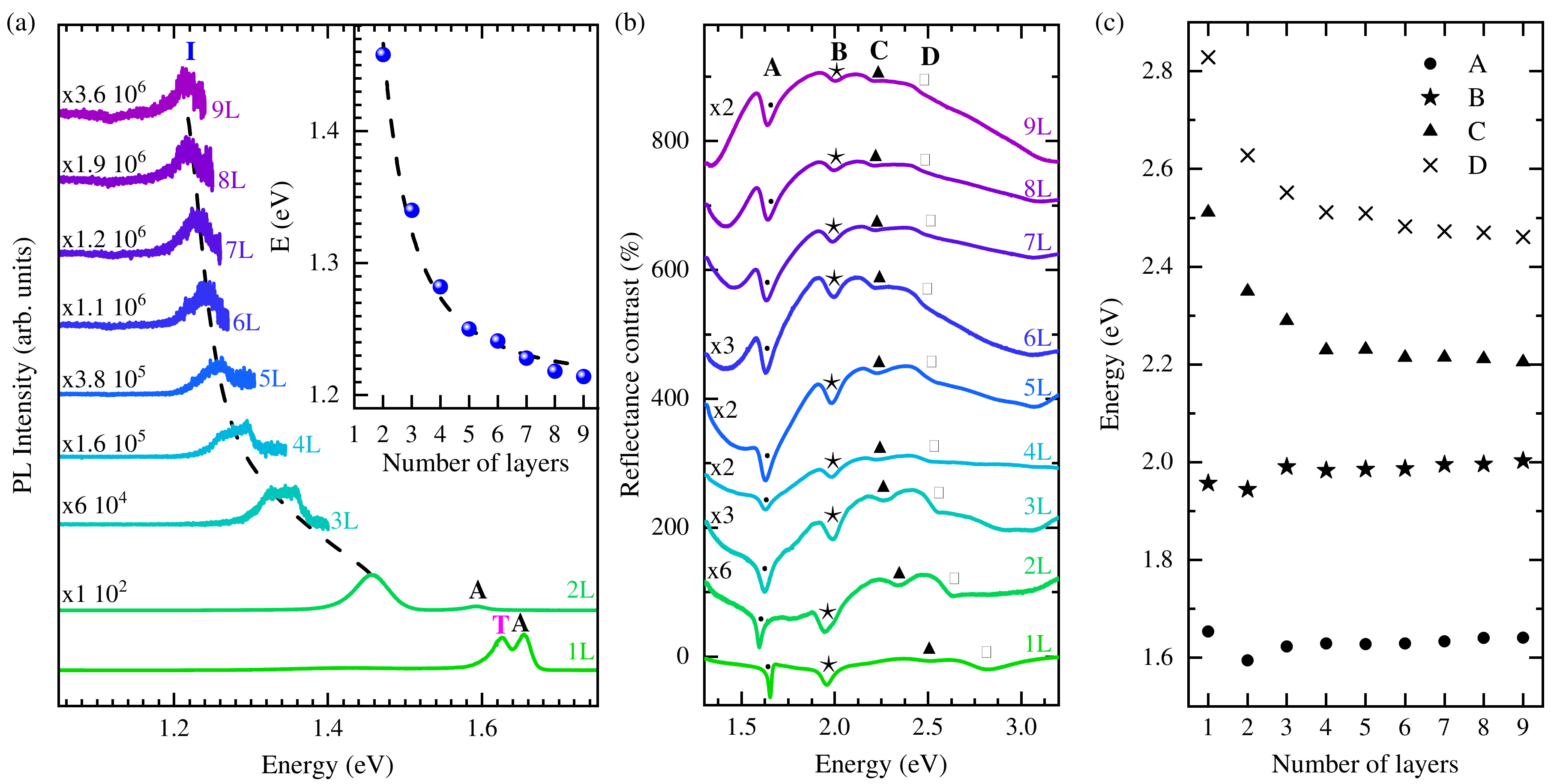}
		\caption{(a) Low-temperature ($T=5$~K) photoluminescence spectra of Mo$_{0.58}$W$_{0.42}$Se$_2$ flakes with thicknesses ranging from 1L to 9L, excited with a 2.41~eV laser. The excitation powers were 5~$\mu$W for 1--2L, 100~$\mu$W for 3L, 200~$\mu$W for 4--5L, and 500~$\mu$W for 6--9L. The spectra are normalized to the intensity of the A-exciton emission in the monolayer and vertically offset for clarity. The inset shows the thickness dependence of the indirect-emission energy. The dashed line represents a fit to the quantum-confinement model described by Eq.~(\ref{eq:quantum_confinement}). (b) Reflectance contrast spectra of Mo$_{0.58}$W$_{0.42}$Se$_2$ flakes as a function of layer thickness. (c) Thickness dependence of the energies of the optical resonances identified in the reflectance contrast spectra.}
		\label{fig:PL_and_RC}
	\end{figure*}

	To quantitatively describe the thickness evolution of the shear-mode energies, we employ the linear-chain model, which has been successfully applied to rigid-layer vibrations in numerous layered TMDs, including MoTe$_2$~\cite{Froehlicher2015,Grzeszczyk2016}, MoS$_2$~\cite{Song2016}, MoSe$_2$~\cite{OBrien2015},  WSe$_2$~\cite{OBrien2015}, and WS$_2$~\cite{Zhao2013}. 
	Within this model, each Se--(Mo/W)--Se layer is treated as a rigid point mass characterized by its areal mass density $\mu$, while the interlayer van der Waals interaction is represented by identical massless springs connecting adjacent layers, with an in-plane force constant per unit area $K_s$. 
	For Mo$_{0.58}$W$_{0.42}$Se$_2$, the effective areal mass density is approximated by the composition-weighted average $\mu = 0.58~\mu_{\mathrm{MoSe_2}} + 0.42~ \mu_{\mathrm{WSe_2}}$, where $\mu_{\mathrm{MoSe_2}} = 4.4\times10^{-6}~\mathrm{kg~m^{-2}}$ and \mbox{$\mu_{\mathrm{WSe_2}} = 5.9\times10^{-6}~\mathrm{kg~m^{-2}}$} are the areal mass densities of 1L MoSe$_2$ and WSe$_2$, respectively~\cite{Li2023}. 
	This gives an effective areal mass density of $\mu = 5.0\times10^{-6}~\mathrm{kg~m^{-2}}$.
	Within the linear-chain model, the Raman shift of the shear mode is given by
	\begin{equation}
		\omega_\alpha=
		\sqrt{\frac{K_s}{2\pi^2\mu c^2}}
		\sqrt{1-\cos\left(\frac{(\alpha-1)\pi}{N}\right)},
		\label{eq:lcm}
	\end{equation}
	where $c$ is the speed of light in vacuum, $N$ is the number of layers, and $\alpha$ denotes the shear-mode branch index. The experimentally observed branches correspond to $\alpha=N$ and $\alpha=N-2$. A simultaneous fit of the model to both branches, shown by the solid lines in Fig.~\ref{fig:RS}(d), accurately reproduces their evolution over the investigated thickness range and yields an interlayer force constant per unit area of
	$K_s=(2.996\pm0.015)\times10^{19}~\mathrm{N~m^{-3}}$.
	This value lies between those reported for the parent compounds MoSe$_2$ and WSe$_2$, namely $2.92\times10^{19}$ and $3.07\times10^{19}~\mathrm{N~m^{-3}}$, respectively~\cite{Zhao2013,Kim2016,Sriv2018}, and is also comparable to the corresponding values for MoS$_2$ and WS$_2$, $2.72\times10^{19}$ and $2.99\times10^{19}~\mathrm{N~m^{-3}}$, respectively~\cite{Golasa2016_acta,Yang2017,Sriv2018}. 
	The intermediate value obtained for Mo$_{0.58}$W$_{0.42}$Se$_2$ indicates that alloying only moderately modifies the interlayer interaction, which remains close to that of the binary parent compounds. 
	This is consistent with the similar interatomic force constants of MoSe$_2$ and WSe$_2$, suggesting that alloying primarily produces quantitative shifts in the phonon energies rather than a qualitative reconstruction of the phonon spectrum~\cite{Proupin2024}.
	
	
	\subsection{Thickness-dependent photoluminescence and reflectance contrast spectra}
	
	To investigate the evolution of the electronic band structure with layer thickness, we performed low-temperature PL measurements at $T=5$~K on Mo$_{0.58}$W$_{0.42}$Se$_2$ flakes, as shown in Fig.~\ref{fig:PL_and_RC}(a). 
	As expected for Mo- and W-based TMDs~\cite{Mak2010,Splendiani2010,Gutierrez2012,Arora2015W,Arora2015Mo,Molas2017Nanoscale}, the 1L exhibits a substantially stronger PL response than the multilayer flakes. 
	The emission intensity decreases by approximately six orders of magnitude from 1L to 9L, reflecting the crossover from a direct band gap in the monolayer to an indirect band gap in multilayer Mo$_{0.58}$W$_{0.42}$Se$_2$.
	
	The PL spectrum of 1L Mo$_{0.58}$W$_{0.42}$Se$_2$ is dominated by emission in the 1.6--1.7~eV range, associated with direct optical transitions at the $K^\pm$ valleys of the BZ. 
	Two distinct emission lines are resolved within this energy range. 
	The higher-energy feature is assigned to the neutral A exciton, while the lower-energy peak is attributed to the charged exciton (T). 
	The simultaneous observation of both excitonic complexes is consistent with previous reports on 1L Mo$_x$W$_{1-x}$Se$_2$ alloys~\cite{Yang2017,OlkowskaPucko2026}. Their emission energies lie between the corresponding values reported for the parent compounds 1L MoSe$_2$ and WSe$_2$~\cite{Tonndorf2013,Arora2015Mo,Arora2015W,Xie2015}.

	Signatures of the direct-gap emission remain visible in the 2L spectra. Starting from 2L, the PL spectra are dominated by a broad lower-energy emission band, denoted as I, which is assigned to a phonon-assisted indirect recombination process. 
	By analogy with other multilayer Mo- and W-based TMDs, this emission is attributed to the recombination of holes near the valence-band (VB) maximum at the $\Gamma$ point with electrons occupying the conduction-band (CB) minimum at the $\Lambda$ (or Q) valley along the $\Gamma$--K direction~\cite{Molas2017Nanoscale,Kopaczek2021}. Since the initial and final electronic states are located at different points of the BZ, phonon participation is required to conserve crystal momentum during radiative recombination~\cite{Molas2017Nanoscale}.
	The thickness dependence of the indirect-transition energy is summarized in the inset of Fig.~\ref{fig:PL_and_RC}(a). 
	Its energy decreases systematically from approximately 1.46~eV in 2L to about 1.21~eV in 9L. 
	This pronounced redshift reflects the progressive evolution of the electronic band structure toward the bulk limit and is characteristic of multilayer TMDs~\cite{Padilha2014}.
	
	\begin{table*}[t]
		\centering
		\caption{Comparison of selected vibrational and optical parameters of MoSe$_2$, Mo$_{0.58}$W$_{0.42}$Se$_2$, and WSe$_2$.}
		\label{tab:comparison}
		\begin{tabular}{lccc}
			\hline
			Parameter 
			& MoSe$_2$ 
			& Mo$_{0.58}$W$_{0.42}$Se$_2$ 
			& WSe$_2$ \\
			\hline
			A$_{1g}$ energy (1L) (cm$^{-1}$) 
			& 242 \cite{Zhang2019_2} & 244.60$\pm$0.75 & 249.5 \cite{Zhao2013} \\
			A$_{1g}$ shift (cm$^{-1}$) 
			& 1.5 \cite{Zhang2019_2} & 2.38$\pm$0.92 & 1.4 \cite{Zhao2013}\\
			Shear-mode energy (2L) (cm$^{-1}$) 
			& 19~\cite{OBrien2015}& 17.87$\pm$0.39 & 17~\cite{OBrien2015} \\ 
			Interlayer force constant
			$K_s$ ($10^{19}$ N\,m$^{-3}$) 
			& 2.92~\cite{Sriv2018} & 2.996$\pm$0.015 & 3.07 \cite{Sriv2018} \\  
			A-exciton energy (1L) (eV) 
			& 1.620$\pm$0.008\cite{Liu2020} & 1.653 & 1.710$\pm$0.009\cite{Liu2020} \\
			B-exciton energy (1L) (eV) 
			& 1.860$\pm$0.009\cite{Liu2020} & 1.957 & 2.160$\pm$0.011\cite{Liu2020} \\
			C-exciton energy (1L) (eV) 
			& 2.52\cite{Kopaczek2022} & 2.511 & 2.45\cite{Kopaczek2022} \\
			D-exciton energy (1L) (eV) 
			& - & 2.828 & 2.78\cite{Kopaczek2022} \\
			Indirect-transition energy
			$E_0$ (eV) 
			& 1.164$\pm$0.020~\cite{Kopaczek2022} & 1.203 & 1.227$\pm$0.020~\cite{Kopaczek2022} \\
			\hline
		\end{tabular}
	\end{table*}
	
	A comprehensive theoretical description of the observed thickness dependence would require a detailed analysis of the electronic band structure. 
	Nevertheless, the experimental data are well reproduced by a simple quantum-confinement model in which charge carriers are confined within a one-dimensional rectangular potential well with infinite barriers~\cite{Molas2017Nanoscale}. 
	The thickness dependence of the indirect-transition energy, shown by the dashed line in the inset of Fig.~\ref{fig:PL_and_RC}(a), is described by
	\begin{equation}
		\label{eq:quantum_confinement}
		E(N)=E_0+\frac{\alpha}{N^2}
		=E_0+\frac{\pi^2\hbar^2}{2\mu_\perp L^2},
	\end{equation}
	where $N$ is the number of layers, $E_0$ is the indirect-transition energy in the bulk limit, $\alpha$ is the quantum-confinement parameter, $\hbar$ is the reduced Planck constant, and $L=Na$ is the flake thickness. 
	Here, $a=0.7~\mathrm{nm}$ is taken as the thickness of a single Mo$_{0.58}$W$_{0.42}$Se$_2$ layer~\cite{Tongay2014}. 
	The parameter $\mu_\perp=\left(1/m_e+1/m_h\right)^{-1}$ represents the out-of-plane reduced effective mass, where $m_e$ and $m_h$ are the effective masses associated with the parabolic out-of-plane dispersions of electrons at the CB minimum in the $\Lambda$ valley and holes at the VB maximum at the $\Gamma$ point, respectively.
	
	The fit yields $E_0=1.203~\mathrm{eV}$ and $\alpha=1.028~\mathrm{eV}$, corresponding to an out-of-plane reduced effective mass of $\mu_\perp=0.75\,m_0$, where $m_0$ is the free-electron mass. 
	The extrapolated value of $E_0=1.203~\mathrm{eV}$ can therefore be regarded as the indirect-transition energy in the bulk limit of Mo$_{0.58}$W$_{0.42}$Se$_2$ within the framework of the adopted quantum-confinement model. This value is in very good agreement with the indirect band-gap exciton of 1.198 eV reported for bulk Mo$_{0.5}$W$_{0.5}$Se$_2$ at low temperature\cite{Kam_1984}. 

	
	To further investigate the evolution of the electronic band structure with layer thickness, we performed low-temperature ($T=5$~K) RC measurements on the same Mo$_{0.58}$W$_{0.42}$Se$_2$ flakes (1L--9L) investigated by PL, as shown in Fig.~\ref{fig:PL_and_RC}(b).
	Four pronounced resonances, labeled A, B, C, and D, are identified in the RC spectra, following the nomenclature commonly adopted for Mo$_{1-x}$W$_x$Se$_2$ alloys~\cite{Kopaczek2021,Taank2024}.
	The A and B resonances are assigned to direct excitonic transitions at the $K^\pm$ valleys of the BZ, with their energy separation originating from the spin--orbit splitting of the VB~\cite{Zhu2011,Molina-Sanchez2013,Sun2013,Kormanyos2015,Koperski2017,Wang2018}.
	In contrast, the higher-energy C and D resonances are associated with optical transitions in the band-nesting region along the $\Gamma$--$K^\pm$ direction, where the VB and CB become nearly parallel, giving rise to a large joint density of states.
	Specifically, the C resonance originates from transitions between the upper VB and the lowest CB, whereas the D resonance involves transitions between the lower VB (VB-1) and the lowest CB~\cite{Kopaczek2021}.

	The thickness dependence of the resonance energies is summarized in Fig.~\ref{fig:PL_and_RC}(c).
	The A and B resonances exhibit only a weak dependence on layer thickness, in agreement with the expected behavior of band-edge excitons in TMDs~\cite{Wang2015,Koperski2017,arora2017}. 
	This behavior can be understood as resulting from an approximate compensation between the reduction in exciton binding energy and the renormalization of the quasiparticle band gap with increasing layer thickness.
	Because the A and B excitons originate from electronic states at the $K^\pm$ valleys, which are largely localized within individual layers, the enhanced dielectric screening modifies both quantities in a similar manner, leaving the optical transition energies nearly unchanged~\cite{Ugeda2014, Koperski2017,arora2017}.
	Small deviations observed for the 1L and 2L most likely reflect changes in the electronic structure in the few-layer limit~\cite{Koperski2017, Molas2017Nanoscale}.
	Importantly, for both the 1L and 2L, the A-resonance energy extracted from the RC spectra agrees well with the A-exciton emission energy determined from the PL measurements, confirming that both techniques probe the same direct excitonic transition\cite{arora2017,Koperski2017}.
	
	In contrast, the C and D resonances exhibit a pronounced redshift with increasing layer number and gradually approach nearly constant energies in thicker flakes. Their much stronger thickness dependence, compared with that of the A and B resonances, indicates that the electronic states involved in the band-nesting transitions are considerably more sensitive to interlayer coupling than the band-edge states at the $K^\pm$ valleys. Increasing the number of layers therefore modifies the corresponding band dispersions and transition energies much more strongly than at the $K^\pm$ valleys.
	This behavior is consistent with previous experimental and theoretical studies of TMDs, which have shown that the electronic structure in the band-nesting region evolves much more strongly with increasing thickness than that at the $K^\pm$ valleys~\cite{Molina-Sanchez2013,Padilha2014,Kopaczek2021,Taank2024}. 
	The high-energy C and D resonances therefore provide a particularly sensitive spectroscopic probe of thickness-induced modifications of the electronic band structure in Mo$_{x}$W$_{1-x}$Se$_2$ alloys.
	Overall, the combined analysis of the A--D resonances provides a comprehensive picture of the evolution of the electronic band structure with increasing layer thickness in Mo$_{0.58}$W$_{0.42}$Se$_2$.
	
	To place the obtained results in the context of the parent compounds, Table~\ref{tab:comparison} summarizes selected vibrational and optical parameters of Mo$_{0.58}$W$_{0.42}$Se$_2$ together with the corresponding values reported for MoSe$_2$ and WSe$_2$. The alloy generally exhibits intermediate values between those of the two parent compounds, while preserving the characteristic thickness-dependent trends of layered TMDs.
	
	\section{Summary}
	
	In conclusion, we have systematically investigated the thickness-dependent optical and vibrational properties of mechanically exfoliated Mo$_{0.58}$W$_{0.42}$Se$_2$ flakes using RS, PL, and RC spectroscopy, supported by first-principles phonon calculations. 
	The phonon dispersion enabled the assignment of both first- and second-order Raman features, while the thickness evolution of the high-frequency A$_{\mathrm{1g}}$ phonon and the low-frequency interlayer shear mode provided sensitive fingerprints of layer number and interlayer coupling. 
	PL measurements revealed the crossover from a direct band gap in the monolayer to an indirect band gap in multilayers, with the thickness dependence of the indirect transition well described by a quantum-confinement model. 
	RC spectroscopy further showed that the A and B excitonic resonances remain nearly thickness independent, whereas the high-energy C and D resonances exhibit a pronounced redshift, reflecting substantial modifications of the electronic structure in the band-nesting region.
	Overall, our results provide a comprehensive picture of the evolution of lattice dynamics and electronic structure with layer thickness and establish reliable spectroscopic fingerprints for the characterization of Mo$_{x}$W$_{1-x}$Se$_2$ alloys.


	\section{Methods}
	\subsection{Samples fabrication}
	Thin Mo$_{0.58}$W$_{0.42}$Se$_2$ flakes were mechanically exfoliated from a bulk crystal (HQ Graphene) onto Si\textbackslash SiO$_2$ substrates with a 285-nm-thick SiO$_2$ layer using a polydimethylsiloxane (PDMS)-assisted exfoliation technique.
	The exfoliated flakes were initially identified by optical microscopy based on their optical contrast, and their thicknesses were subsequently determined by AFM.
	
	The \ce{Mo_{0.58}W_{0.42}Se_2} alloy was examined by energy dispersive X-ray analysis (EDX) using a Carl Zeiss Sigma HV field-emission scanning electron microscope (FE-SEM) equipped with a Gemini electron column, an InLens secondary electron detector, a backscattered electron (BSE) detector, and a Bruker Quantax XFlash 6|10 energy-dispersive X-ray spectroscopy (EDX) detector.

	\subsection{Experimental setup}
	Room-temperature Raman scattering and low-temperature ($T=5$~K) PL measurements were performed using the 514.5~nm (2.41~eV) line of a continuous-wave diode laser. The excitation power was fixed at approximately 500~$\mu$W on the sample for the Raman measurements, whereas it was varied between 5 and 500~$\mu$W for the PL measurements.
	The excitation beam was focused onto the sample through a $50\times$ long-working-distance objective (NA = 0.55), producing a laser spot approximately 1~$\mu$m in diameter. 
	The scattered light was collected by the same objective, spectrally filtered using long-pass or volume Bragg grating notch filters, dispersed by a 0.75~m spectrometer equipped with 300 or 1800~grooves/mm diffraction gratings, and detected using a liquid-nitrogen-cooled charge-coupled device (CCD) camera.
	For the low-temperature measurements, the samples were mounted on the cold finger of a continuous-flow helium cryostat. 
	To enable low-frequency Raman measurements down to approximately 5~cm$^{-1}$ from the laser line, a set of Bragg filters was implemented in both the excitation and detection paths. 
	Reflectance-contrast measurements were performed using the same optical setup as for the PL measurements. 
	In this case, the laser excitation was replaced by a stabilized halogen lamp serving as a broadband white-light source. 
	The reflectance contrast was calculated according to Ref.~\cite{Arora2015W}: $\mathrm{RC}=
	\frac{R_{\mathrm{flk}}-R_{\mathrm{sub}}}
	{R_{\mathrm{flk}}+R_{\mathrm{sub}}}\times 100\%,
	\label{4}
	$ where $R_{\mathrm{flk}}$ and $R_{\mathrm{sub}}$ are the reflectance spectra acquired from the flake and the substrate, respectively.
	
	\subsection{Density functional theory calculations}
	Density functional theory calculations were performed in Quantum Espresso \cite{10.1063/5.0005082,Giannozzi_2017,Giannozzi_2009} with scalar-relativistic optimized norm-conserving Vanderbilt pseudopotentials SG15v1.2 \cite{SCHLIPF201536}. The exchange--correlation energy was described using the Perdew--Burke--Ernzerhof parametrization of the generalized gradients approximation (GGA-PBE) \cite{Perdew1996}. Grimme's D3 correction was applied to account for the vdW interactions \cite{D3}. The kinetic energy cutoff for wavefunction was set to 90~Ry, and for charge density and potential to 360~Ry. The first Brillouin zone was sampled with a $12\times12\times4$ $\Gamma$-centered Monkhorst--Pack k-points grid. The atomic positions were optimized until the residual forces on each atom were below $10^{-5}$~Ry/Bohr. Lattice parameters were set to $a=3.285$~\AA  and $c=12.93$~\AA, as averaged between the values for parent crystals \cite{MoSe2,WSe2}.
	
	The phonon dispersion of bulk Mo$_{0.5}$W$_{0.5}$Se$_2$ was calculated using density functional perturbation theory on a $6\times6\times2$ q-points grid, as implemented in Quantum Espresso. An ordered Mo$_{0.5}$W$_{0.5}$Se$_2$ structure was constructed in the primitive cell, in which Mo and W atoms periodically occupy the transition-metal sublattices. 
	
	\section{Acknowledgments}
	The work was supported by the National Science Centre, Poland (grant no. 2025/57/B/ST5/03288). DFT calculations have been carried out in Wroclaw Centre for Networking and Supercomputing.
	
	\section{Author Contributions}
	M.R.M. and K.O.-P. conceived and supervised the project.
	S.S. and K.O.-P. prepared the Mo$_{0.58}$W$_{0.42}$Se$_2$ samples by mechanical exfoliation. 
	T.W. performed the first-principles phonon calculations. 
	S.S., K.O.-P. and G.K. carried out the RS, PL, and RC measurements. 
	M.B. and P.W. performed the SEM--EDX measurements. 
	S.S., K.O.-P., M.R.M., and A.B. analyzed the experimental data. 
	E.B. and A.P. contributed to the interpretation and discussion of the results. 
	S.S, M.R.M, and K.O.-P. wrote the manuscript with input from all authors. 
	All authors discussed the results and approved the final version of the manuscript.
	
	\section{Conflicts of interests}
	There are no conflicts to declare.
	
	\section {Availability of data and materials}
	The datasets generated and analysed during the current study are publicly available at the following link:

	\bibliographystyle{apsrev4-2}
	\bibliography{biblio.bib}
	
	\newpage
	\onecolumngrid

	\renewcommand{\thefigure}{S\arabic{section}.\arabic{figure}}
\renewcommand{\thesection}{S\arabic{section}}

	\begin{center}
		
		{\large{\bf Supplemental Material: \\
				Layer-Dependent Vibrational and Optical Properties of
				\ce{Mo_{0.58}W_{0.42}Se_2} Alloy}}
		
		\vskip0.5\baselineskip
		
		Szymon Socha,$^{1}$
		Tomasz Woźniak,$^{1,2}$
		Elena Blundo,$^{3}$
		Małgorzata Brzoska,$^{1}$
		Grzegorz Krasucki,$^{1}$
		Piotr Wróbel,$^{1}$
		Antonio Polimeni,$^{3}$
		Adam Babiński,$^{1}$
		Maciej R. Molas,$^{1}$
		and Katarzyna Olkowska-Pucko$^{1}$
		
		\vskip0.5\baselineskip
		
		{\em
			$^{1}$Faculty of Physics, University of Warsaw,
			02-093 Warsaw, Poland\\
			$^{2}$Institute of Theoretical Physics,
			Wrocław University of Science and Technology,
			50-370 Wrocław, Poland\\
			$^{3}$Physics Department, Sapienza University of Rome,
			00185 Rome, Italy
		}
		
	\end{center}
	
	\vskip0.5\baselineskip
	
	This Supplemental Material provides additional information on the
	elemental and structural characterization of the investigated
	Mo$_{0.58}$W$_{0.42}$Se$_2$ flakes and on the assignment of the
	observed Raman features. It includes EDX analysis of the bulk crystal,
	AFM characterization of the layer thicknesses, comparison of the
	experimental Raman shifts with the calculated phonon dispersion, and
	the calculated phonon density of states.

	\setcounter{section}{0}
	\setcounter{figure}{0}
	\section{EDX analysis of the bulk MoWSe$_2$ crystal
		\label{SI:EDX}}
	
\begin{figure}[htbp]
		\centering
		\includegraphics[width=1\linewidth]{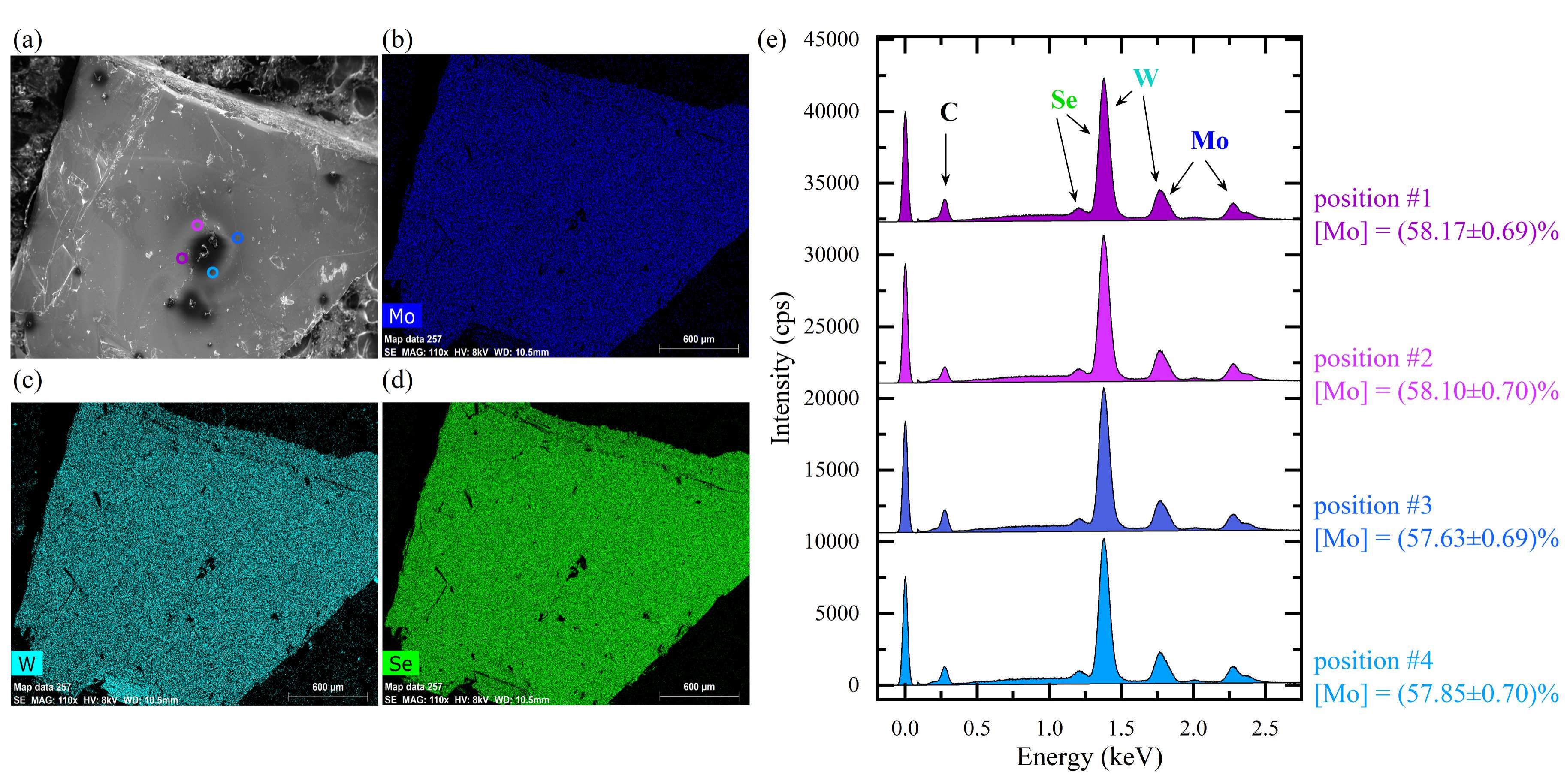}
		\caption{SEM-EDX map of around 2.5 mm big piece of
			Mo$_x$W$_{1-x}$Se$_2$. (a) shows the electron image of the crystal,
			while (b-d) show the elemental mapping analysis of Mo, W, and Se.
			(e) EDX high-resolution spectra acquired at four different positions
			(highlighted by coloured circles in the SEM image) over the
			MoWSe$_2$ crystal. The spectra are stacked by a constant
			y-offset of 10000 counts for ease of comparison. From a quantitative
			analysis of the Mo and W lines, we estimated the relative Mo
			composition ($x$) displayed on the right.}
		\label{fig:EDX}
	\end{figure}

	The Mo$_x$W$_{1-x}$Se$_2$ crystal used in this work was characterized
	by a nominal composition $x = 0.5$. To verify the real composition of
	the sample and its uniformity over large (hundreds of $\mu$m) scales,
	we performed scanning electron microscopy (SEM) with energy dispersive
	X-ray analysis (EDX) (see Methods for details). Figure
	\ref{fig:EDX}~(a) shows the SEM image (top-left panel) of a 2.5 mm size
	crystal of Mo$_x$W$_{1-x}$Se$_2$ along with the elemental mapping
	analysis of Mo [Fig.~\ref{fig:EDX}~(b)], W [Fig.~\ref{fig:EDX}~(c)],
	and Se [Fig.~\ref{fig:EDX}~(d)]. Indeed, a quite uniform intensity is
	found over the whole crystal, demonstrating that the alloys grow
	uniformly across large scales and that no agglomerates of variable
	composition are found.
	
	To have precise information on the sample composition, we took
	highly-resolved EDX spectra at four different points of the crystal,
	as shown in Fig.~\ref{fig:EDX}~(e). From a quantitative elemental
	analysis, we derived the $x$ values displayed on the right. Indeed,
	a relatively small variability is observed, with $x$ being about
	58\%.

	
	\section{Atomic force microscopy
		\label{AFM}}
	
	The number of layers in the investigated flakes was initially assigned
	based on the analysis of their photoluminescence (PL) and Raman
	scattering (RS) spectra, as discussed in the main text. To independently
	verify these assignments, atomic force microscopy (AFM) measurements
	were performed for all investigated flakes except the one identified
	as 9L, which degraded before the AFM measurements could be carried out.
	
\begin{figure}[htbp]
		\centering
		\includegraphics[width=1\linewidth]{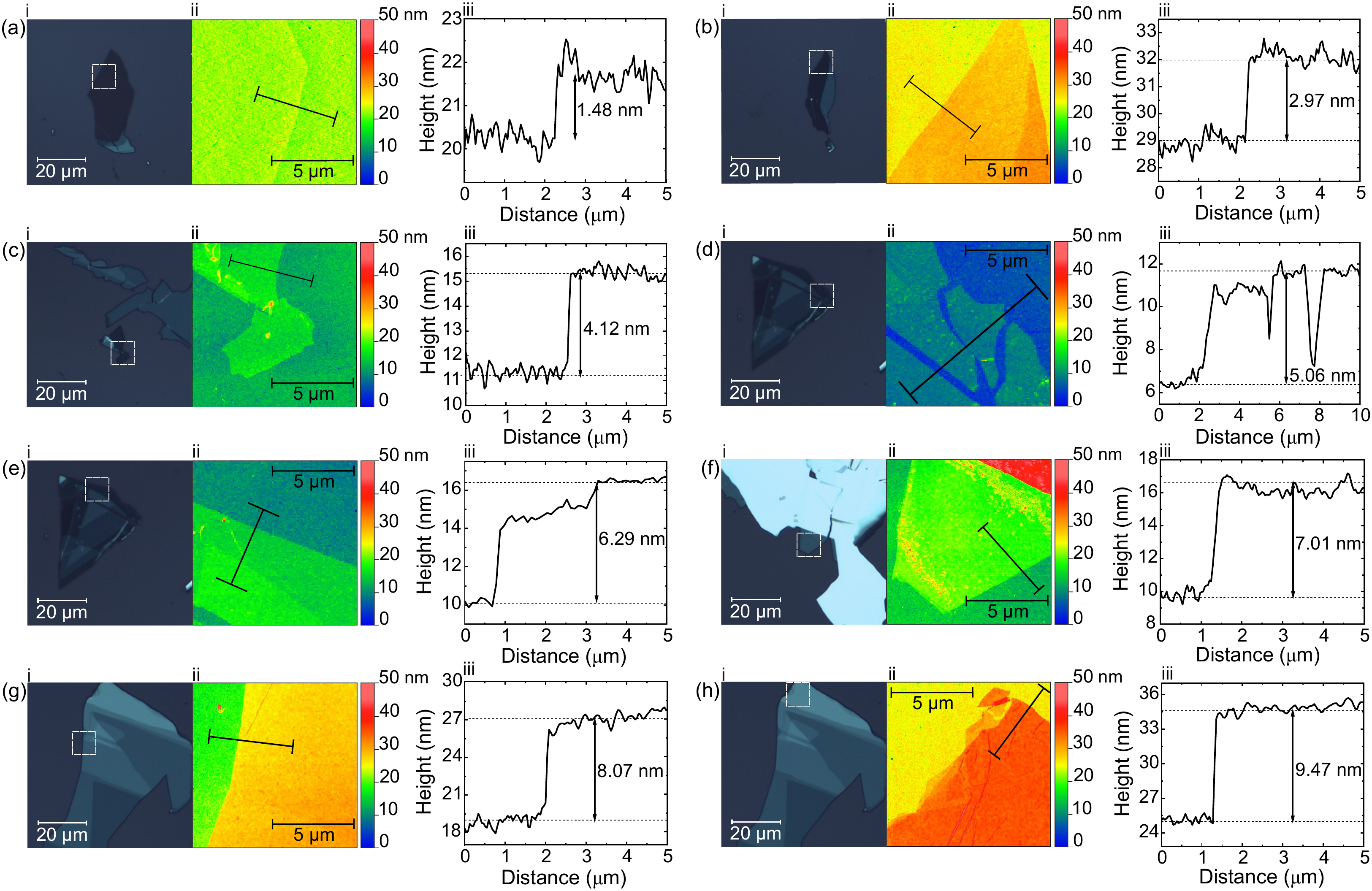}
		\caption{(ai)-(hi) Optical images of the
			Mo$_{0.58}$W$_{0.42}$Se$_2$ flakes under study.
			(aii)-(hii) False-color atomic-force-microscopy images of the
			areas enclosed by the dashed white boxes in (ai)-(hi).
			(aiii)-(hiii) Height profiles measured along the black lines
			crossing the edges of the flakes.}
		\label{fig:AFM}
	\end{figure}

	Optical microscopy images of the investigated flakes with thicknesses
	ranging from 1L to 8L are shown in Fig.~\ref{fig:AFM}(ai)--(hi).
	The regions enclosed by the dashed white squares were subsequently
	scanned by AFM, and the corresponding topography maps are presented
	in Fig.~\ref{fig:AFM}(aii)--(hii). Height profiles extracted along
	the black lines crossing the flake edges are shown in
	Fig.~\ref{fig:AFM}(aiii)--(hiii).
	
	The AFM topography maps exhibit a relatively high level of noise,
	particularly for the thinnest flakes, which could not be completely
	removed by the plane-leveling procedure. This residual noise most
	likely originates from the instrumental resolution of the AFM setup
	and/or contamination remaining on the sample surface after the
	exfoliation process. Nevertheless, the step at the flake edge is
	clearly resolved for all investigated samples, enabling reliable
	determination of the flake thickness.
	
	The flake thickness was determined from the height difference between
	the substrate and the flake surface, as indicated in the height
	profiles shown in Fig.~\ref{fig:AFM}(aiii)--(hiii). Reported AFM
	heights of monolayer MoSe$_2$ and WSe$_2$ are typically not smaller
	than approximately 0.80 and 0.77~nm, respectively
	\cite{Kim2016,Xiao2018,Diware2017,Sha2017}. Taking these values into
	account, we adopted 1.53~nm (twice the reported monolayer height of
	WSe$_2$) as a conservative upper limit for identifying a monolayer.
	The flake assigned as 1L exhibits an AFM height of 1.48~nm, fully
	consistent with this criterion. Moreover, this assignment is
	independently confirmed by its pronounced PL intensity, which is
	characteristic of a direct-gap monolayer.

	
	\section{Assignment of Raman modes
		\label{SI:assigment}}
	
	The Raman features observed for the 9L
	Mo$_{0.58}$W$_{0.42}$Se$_2$ flake were assigned by comparing the
	experimental Raman shifts with the phonon energies obtained from the
	DFT-calculated phonon dispersion of bulk Mo$_{0.5}$W$_{0.5}$Se$_2$,
	supported by previous reports for MoSe$_2$, WSe$_2$, and related
	Mo$_x$W$_{1-x}$Se$_2$ alloys
	\cite{Zhang2014,Zhang2015_raman,Shinde2021}. Owing to the resonant
	excitation conditions, the spectra contain both first-order
	Raman-active phonons and higher-order features involving phonons from
	different regions of the Brillouin zone. Table~\ref{tab:Mody}
	summarizes the proposed assignments together with the calculated and
	experimental phonon energies and their differences. Overall, good
	agreement is obtained, with the largest deviation of 9~cm$^{-1}$ for
	the $\ce{E^2_{2g}(\Gamma)}$ mode.
	
	\begin{table}[htbp]
		\centering
		\caption{Identified modes in the RS spectrum, their energies
			obtained from the calculated phonon dispersion
			($\omega_{\mathrm{calc}}$) for bulk
			\ce{Mo_{0.5}W_{0.5}Se_2}, the experimental Raman shifts
			($\omega_{\mathrm{exp}}$) for the 9L
			\ce{Mo_{0.58}W_{0.42}Se_2} flake, and their differences
			$\omega_{\mathrm{calc}}-\omega_{\mathrm{exp}}$.}
		\label{tab:Mody}
		\begin{tabular}{c|ccc}
			Mode &
			$\omega_{\mathrm{calc}}$ (cm$^{-1}$) &
			$\omega_{\mathrm{exp}}$ (cm$^{-1}$) &
			$\omega_{\mathrm{calc}}-\omega_{\mathrm{exp}}$ (cm$^{-1}$)
			\\ \hline
			
			$\ce{S/E^1_{2g}}$ & 28 & 25 & 3 \\
			$\ce{LA}$ & 140 & 145 & -5 \\
			$\ce{E_{1g}(\Gamma)}$ & 172 & 170 & 2 \\
			$\ce{TA}+\ce{ZA}$ & 217 & 218 & -1 \\
			$\ce{A_{1g}(\Gamma)}$ & 247 & 247 & 0 \\
			$\ce{E^2_{2g}(\Gamma)}$ & 285 & 294 & -9 \\
			$\ce{E_{\mathrm{1g}}}+\ce{LA}$ & 315 & 317 & -2 \\
			$\ce{B^2_{\mathrm{2g}}(\Gamma)}$ & 348 & 348 & 0 \\
			$\ce{A_{1g}}+\ce{ZA}$ & 372 & 370 & 2 \\
			$\ce{A_{1g}}+\ce{LA}$ & 392 & 400 & -8 \\
			$\ce{E^2_{2g}}+\ce{LA}$ & 439 & 440 & -1 \\
			$\ce{A_{1g}}+\ce{E^2_{2g}}$ & 541 & 551 & -1 \\
			$\ce{2*E^2_{2g}(\Gamma)}$ & 588 & 586 & -2
		\end{tabular}
	\end{table}

	\newpage

	
	\section{Calculated phonon density of states
		\label{SI:DOS}}
	
	\begin{figure}[htbp]
		\centering
		\includegraphics[width=0.7\linewidth]{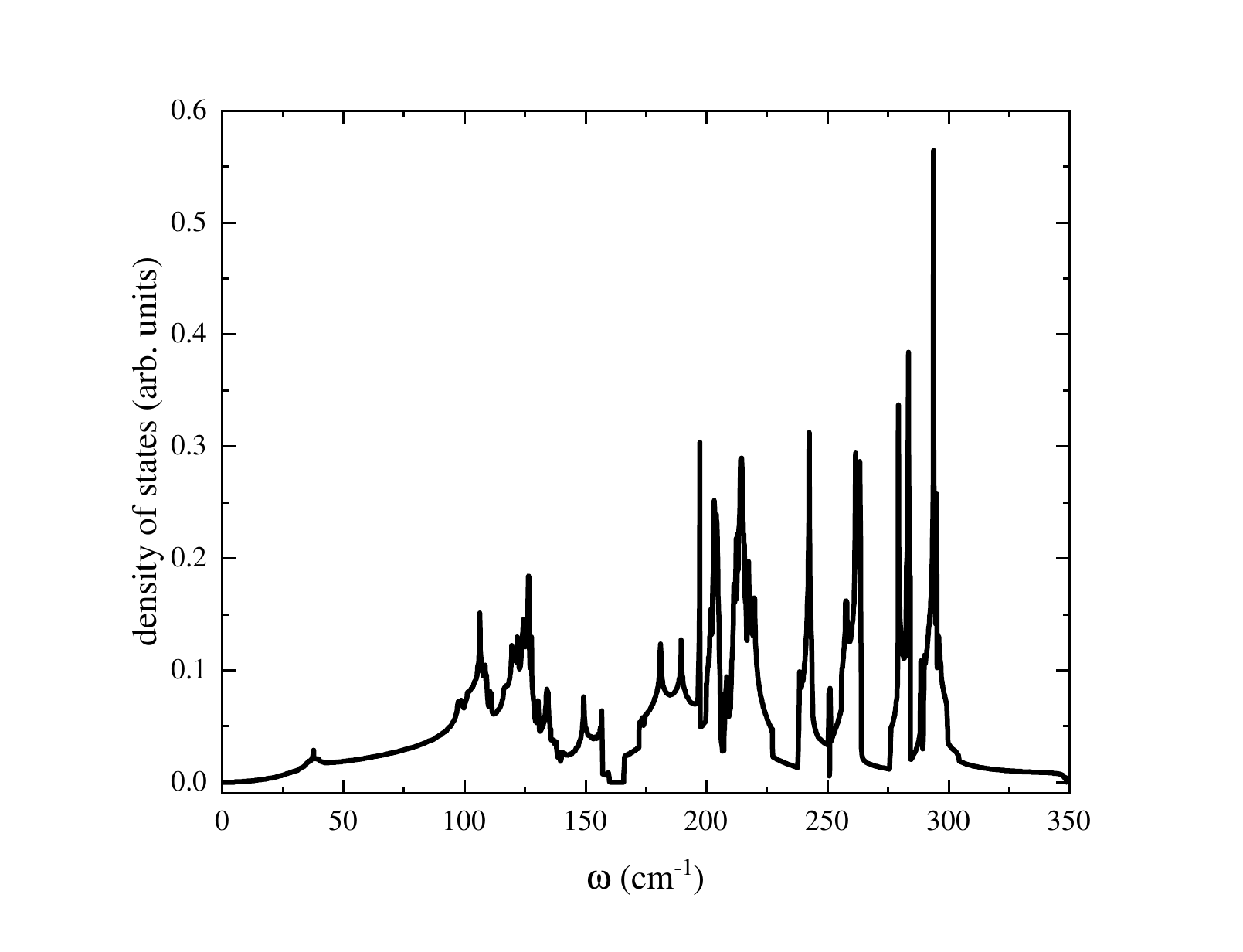}
		\caption{Calculated phonon density of states for bulk
			Mo$_{0.58}$W$_{0.42}$Se$_2$.}
		\label{fig:DOS}
	\end{figure}
	
	Fig.~\ref{fig:DOS} presents the phonon density of states for bulk
	Mo$_{0.58}$W$_{0.42}$Se$_2$, calculated from first principles.
	The spectrum comprises contributions from both acoustic and optical
	phonon modes over the entire frequency range.

\end{document}